\documentclass[sigconf]{acmart}

\setcopyright{none}
\renewcommand\footnotetextcopyrightpermission[1]{}
\usepackage{booktabs}
\usepackage{tabularx}
\usepackage{amsmath}
\usepackage{xspace}
\usepackage{xcolor}
\usepackage{graphicx}

\newcommand{\one}{({\em i}\/)\xspace}
\newcommand{\two}{({\em ii}\/)\xspace}
\newcommand{\three}{({\em iii}\/)\xspace}
\newcommand{\four}{({\em iv}\/)\xspace}

\def\ie{\emph{i.e.,}\xspace}

\usepackage{fancyhdr}
\fancypagestyle{plain}{%
   \fancyhf{} %
   \fancyfoot[L]{under submission to ACM SIGCOMM CCR}%
   \fancyfoot[R]{Volume xx Issue x, 2026}%
}
\fancypagestyle{firstpagestyle}{%
   \fancyhf{} %
   \fancyfoot[L]{under submission to ACM SIGCOMM CCR}%
   \fancyfoot[R]{Volume xx Issue x, 2026}%
}

\acmConference[under submission to ACM SIGCOMM CCR]{}{Volume xx Issue x}{2026}

\newcolumntype{Y}{>{\raggedright\arraybackslash}X}

\begin{document}

\title{Intelligence Delivery Network: Toward an Internet Architecture for the AI Age}


\author{%
Hanling Wang\textsuperscript{1},
Qing Li\textsuperscript{1,*},
Dan Zhao\textsuperscript{1},
Yuhong Song\textsuperscript{1},
Xingchi Chen\textsuperscript{1},
Teng Gao\textsuperscript{1},
Peiyuan Zong\textsuperscript{1},
Zhuyun Qi\textsuperscript{1},
Yue Yu\textsuperscript{1},
Yong Jiang\textsuperscript{2,1}
}

\email{{wanghl03, liq, zhaod01, songyh, chenxch01, gaot, zongpy, qizy, yuy}@pcl.ac.cn, jiangy@sz.tsinghua.edu.cn}

\affiliation{%
  \institution{%
  \textsuperscript{1}Pengcheng Laboratory, Shenzhen, Guangdong, China \\
  \textsuperscript{2}Tsinghua SIGS, Shenzhen, Guangdong, China \\
  \textsuperscript{*}Corresponding author
  }
  \city{}
  \country{}
}

\renewcommand{\shortauthors}{Wang, Li, et al.}

\begin{abstract}

The rapid emergence of AI-powered applications is reshaping the role of the Internet. Users increasingly rely on the network to obtain intelligence services derived from large foundation models, rather than merely to reach remote endpoints or retrieve specific content. Today's dominant deployment paradigm for AI services remains cloud-centric, where user requests are transmitted to remote data centers for centralized inference. Although operationally convenient, this paradigm suffers from latency and jitter, heavy wide-area traffic, limited utilization of distributed heterogeneous compute resources, and growing privacy and governance concerns. In this paper, we propose the \textbf{Intelligence Delivery Network (IDN)}, an Internet architecture that treats AI capabilities as deliverable network services. The key idea is to position, select, reuse, and verify intelligence across cloud, regional, edge, and local environments according to demand locality, resource availability, and policy constraints. We present the system assumptions of IDN, define its core architectural mechanisms, and discuss how capability abstraction, compute resource integration, demand-driven deployment, service routing, state-aware caching, and trust management can jointly support distributed AI services. We believe that IDN provides a practical path toward an Internet architecture for the AI age, making AI capabilities more accessible, efficient, trustworthy, and responsive to diverse application needs.

\end{abstract}

\ccsdesc[500]{Networks~Network architectures}
\ccsdesc[500]{Computer systems organization~Cloud computing}
\ccsdesc[500]{Security and privacy~Distributed systems security}

\keywords{Internet architecture, distributed AI services, edge intelligence, service routing, model serving, caching}

\maketitle

\section{Vision: Intelligence as a Networked Service}

\begin{figure}[t]
  \centering
  \includegraphics[width=0.97\columnwidth]{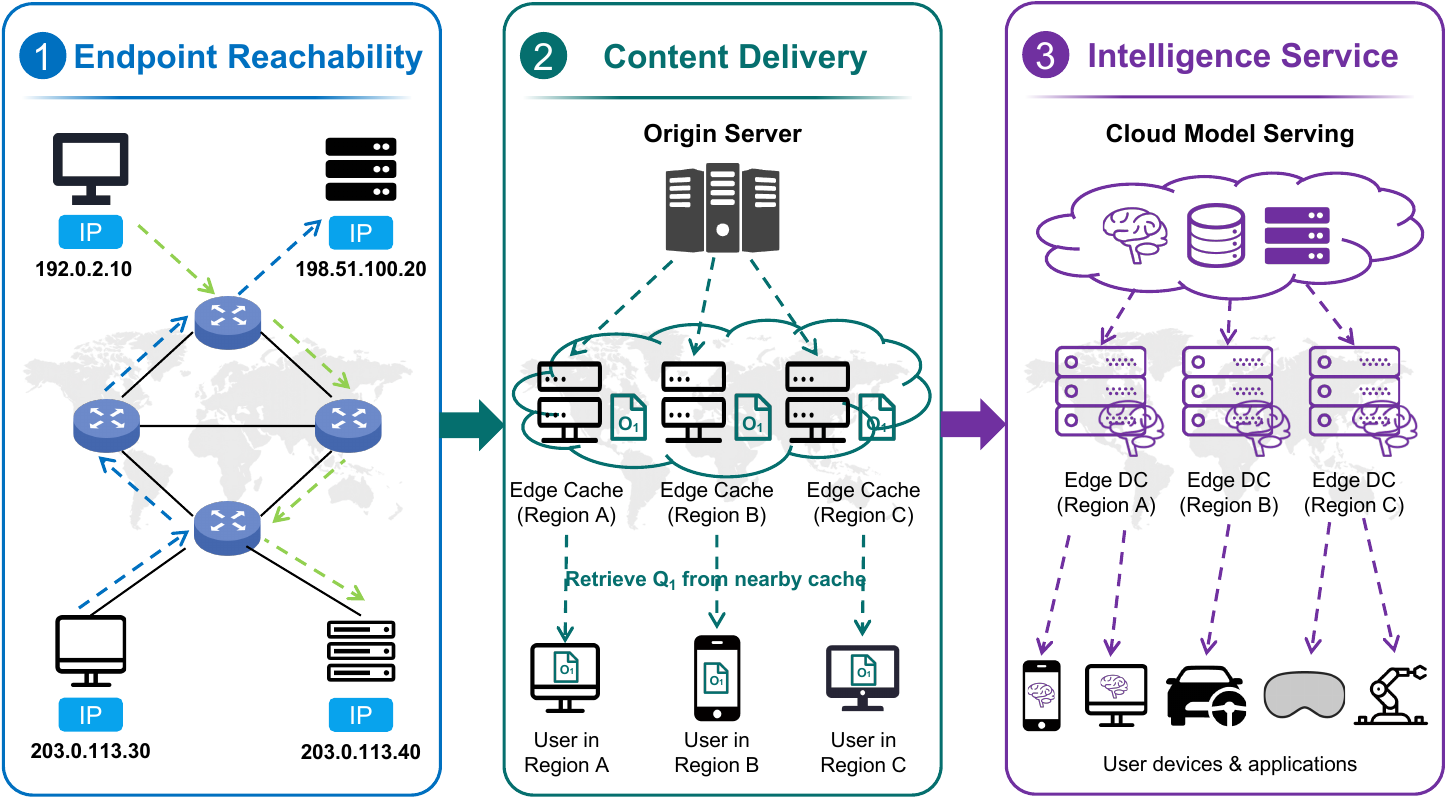}
  \caption{A high-level shift in network abstractions: from endpoint reachability, to content delivery, to intelligence delivery.}
  \Description{The high-level shift of network abstraction from endpoint reachability, to content delivery, to intelligence delivery.}
  \label{fig:paradigm}
\end{figure}

Over the past several decades, the Internet has been repeatedly reshaped by the abstraction it makes common, as shown in Fig. \ref{fig:paradigm}. The early Internet standardized global communication around endpoint addressing and packet delivery, leaving higher-level semantics to applications and end systems~\cite{cerf74,saltzer84,clark88}. Later, web-scale content distribution showed that many workloads are better understood as retrieving and replicating named objects than as maintaining host-to-host conversations. This observation shaped industrial Content Delivery Networks (\textbf{CDNs})~\cite{dilley02,nygren10} and motivated information-centric networking research~\cite{jacobson09,zhang14ndn,xylomenos14}.

A comparable shift is now emerging with the prosperity of large foundation models. Transformer-based Large Language Models (\textbf{LLM}) have enabled general-purpose Artificial Intelligence (\textbf{AI}) capabilities that can be invoked by a variety of applications, including interactive assistants, retrieval-augmented search, code copilots, real-time translation, robotics, and multimodal perception~\cite{lewis20rag,clipper17,orca22,pagedattention23}. In these settings, the central question is no longer simply whether a server is reachable. What matters is whether the system can deliver the appropriate intelligence with acceptable latency, quality, cost, privacy, and operational stability.

Today, however, AI services remain largely cloud-centric. User inputs are sent across the wide-area network to remote data centers, where inference is performed inside provider-controlled serving stacks. Such stacks have made major progress in batching, scheduling, memory management, and model-variant selection~\cite{clipper17,clockwork20,infaas21,orca22,pagedattention23}. Although this approach is operationally attractive, it also reveals several fundamental limitations: \one It introduces nontrivial end-to-end latency and jitter, which are difficult to tolerate in interactive or time-sensitive applications; \two It increases wide-area traffic by repeatedly transporting prompts, retrieved documents, media streams, and other input data to distant locations; \three It leaves many distributed and heterogeneous compute resources at the edge, on premises, and within local networks underutilized. \four It complicates privacy protection, policy compliance, and data governance, since sensitive inputs and execution context may need to cross administrative or geographic boundaries.

These limitations point to the need for a new network architecture designed specifically for delivering intelligence. We observe that distributed AI services resemble content in CDNs in several important respects: demand is often geographically clustered, requests frequently invoke related capabilities, and many applications benefit when execution remains close to users, data sources, or policy boundaries. CDNs improve performance by moving popular objects closer to demand through locality-aware placement, hierarchical coordination, and selective replication~\cite{dilley02,nygren10}. This naturally raises the question: \textit{Can intelligence be delivered in a similar way as CDNs?}

However, AI service may depend on a base model, adapters, tokenization logic, safety filters, retrieval indices, and reusable runtime state such as prompt prefixes or key-value (\textbf{KV}) caches. These elements have compatibility constraints, hardware requirements, loading costs, trust relationships, and service-level implications. As a result, delivering intelligence requires more than object placement. It requires a common way to describe capabilities, expose execution conditions, guide deployment and routing, reuse state, and establish trust in distributed execution.

To this end, we propose the \textbf{Intelligence Delivery Network (IDN)}, an Internet architecture for distributed AI services. IDN is designed to make intelligence available where and when it is needed by enabling the network to position, select, and deliver AI capabilities according to demand locality, resource availability, and policy constraints. In this paper, we present the architectural vision of IDN through six core mechanisms: intelligence capability abstraction, compute resource integration, demand-driven capability deployment, capability-aware service routing, state-aware caching, and secure and trustworthy management. We believe that IDN offers a practical path toward an intelligence-centric Internet and establishes an architectural foundation for delivering intelligence as a network service.

\section{System Assumptions}

\begin{figure}[t]
  \centering
  \includegraphics[width=0.95\linewidth]{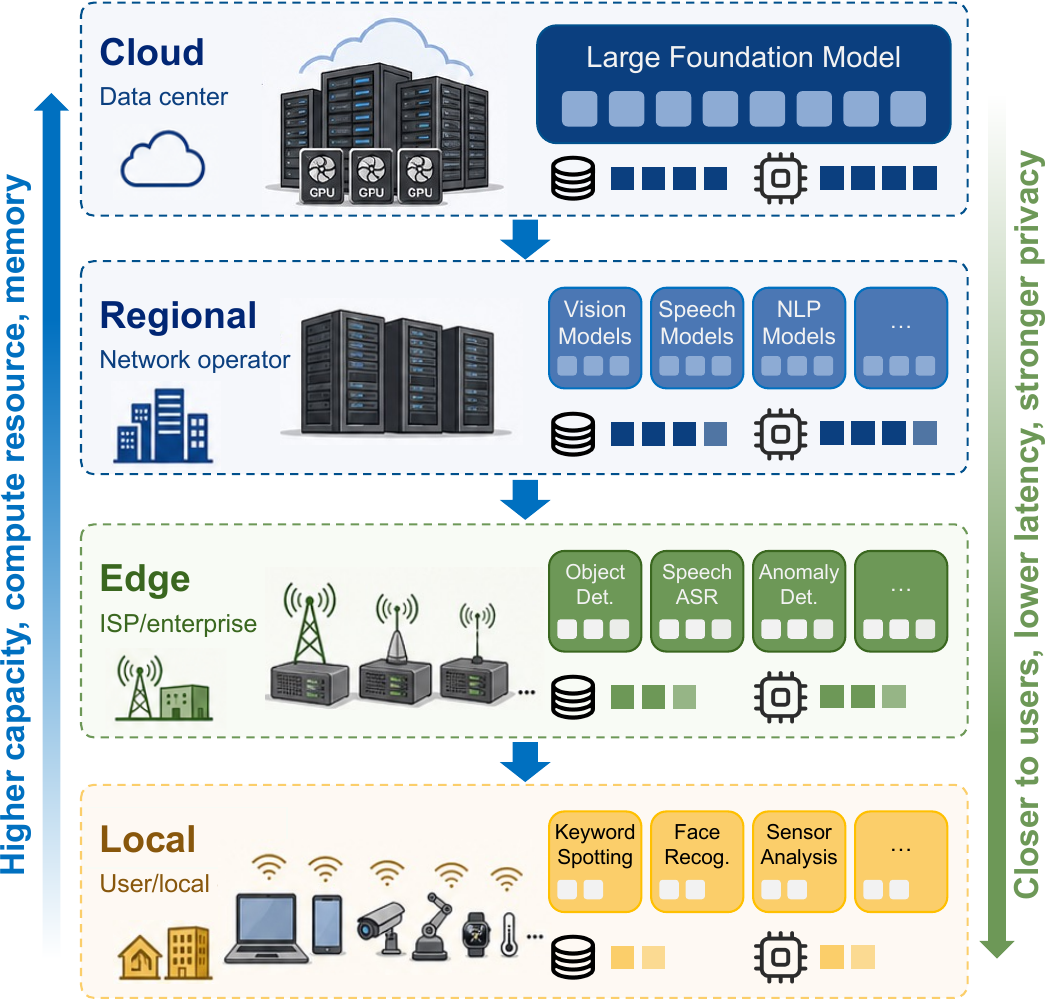}
  \caption{System assumptions of IDN. AI capabilities are delivered over a hierarchy of cloud, regional, edge, and local resources that differ in capacity, proximity, and administrative control.}
  \Description{The figure shows the hierarchical compute resources of cloud, regional, edge, and local.}
  \label{fig:system-model}
\end{figure}

Compute resources in the Internet are inherently hierarchical and heterogeneous, residing at different levels of the network and offering different tradeoffs in capacity, locality, and administrative control~\cite{shi16,satya17,mao17,zhou19}. Modern CDNs already exploit hierarchical deployment for content delivery by replicating popular objects across multiple layers of infrastructure~\cite{dilley02,nygren10}. IDN considers an analogous setting for AI services, \ie intelligence should not be served only from distant cloud clusters, but positioned across multiple levels of the network according to resource availability, proximity to demand, and policy constraints, as shown in Fig. \ref{fig:system-model}.

\subsection{Hierarchical and Heterogeneous Deployment}

We consider an Internet-scale service environment in which AI capabilities are deployed over a hierarchy of resource domains. Cloud data centers host the largest foundation models, maintain the most complete set of capabilities, and provide abundant compute, memory, and storage. Regional or metro-scale nodes offer more limited but still substantial resources and are suitable for hosting frequently used capabilities within a geographic or administrative region. Edge and local nodes operate under tighter resource constraints but can provide lower latency, better locality, and stronger alignment with local privacy or policy requirements. In some cases, enterprise premises or end devices may also contribute limited capabilities when local execution is required.

These locations differ not only in scale, but also in accelerator type, memory capacity, startup overhead, connectivity, reliability, and distance to demand. As a result, intelligence cannot be treated as a uniform deployment object. A monolithic model that fits naturally in a cloud cluster may be infeasible to place at the edge, while a compact task-specialized capability may be more efficient near users than repeatedly invoking a remote foundation model. Similar to CDN, the goal of IDN is not to replicate everything everywhere, but to place the right capability at the right level of the hierarchy.

\begin{figure*}[t]
  \centering
  \includegraphics[width=0.65\textwidth]{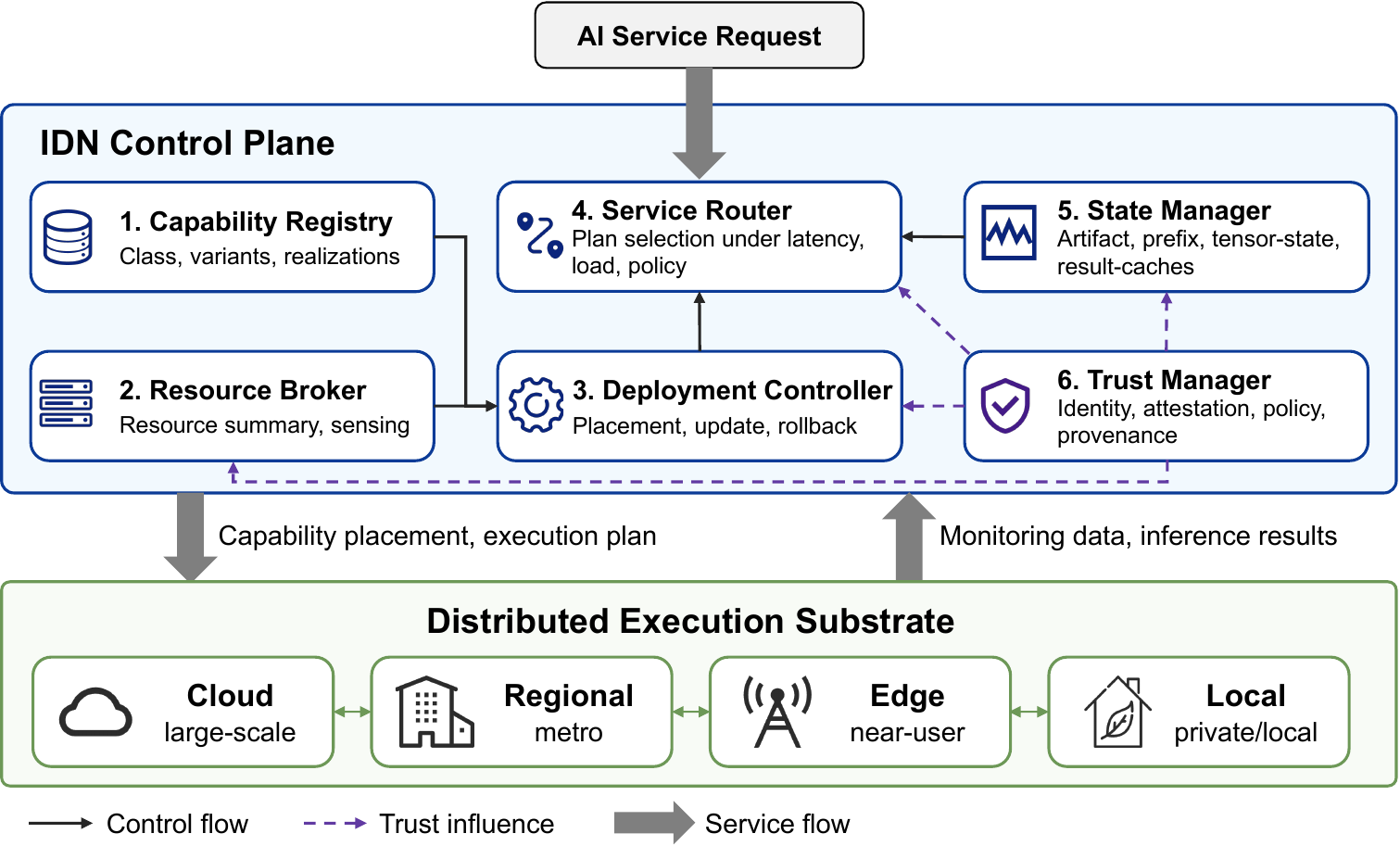}
  \caption{IDN architecture. The six components form a system for describing, placing, routing, reusing, and securing intelligence across heterogeneous resources.}
  \Description{IDN architecture. The six components form a system for describing, placing, routing, reusing, and securing intelligence across heterogeneous resources.}
  \label{fig:idn_architecture}
\end{figure*}

\subsection{Administrative Domains and Control Boundaries}

The deployment environment considered by IDN is also multi-domain. Cloud infrastructure may belong to model providers, regional sites may be operated by service providers or large network operators, edge resources may be managed by ISPs, enterprises, or campus networks, and local resources may be controlled by users with their own policies. Consequently, IDN cannot assume a single global controller with complete authority over all resources.

Instead, each domain may authenticate nodes, admit deployment objects, expose resource conditions, and enforce local security or compliance rules under its own control. Cross-domain intelligence delivery therefore requires explicit capability descriptions, common deployment interfaces, policy-aware request steering, and verifiable execution evidence rather than implicit trust in a provider-local serving stack. This assumption aligns with current computing-aware traffic steering works~\cite{catsFramework,catsUsecases}.

\subsection{Service Model}

We model an intelligence request as a demand for a class of AI capability under performance and policy constraints. A request may specify, explicitly or implicitly, the desired task, quality target, response-time requirement, privacy or locality constraint, and affinity to reusable state. The purpose of IDN is not merely to identify a reachable endpoint, but to determine where and how this requested capability should be realized within distributed infrastructure.

A request may be served in several ways. A nearby node may host a suitable lightweight capability, a regional node may provide a stronger variant with acceptable latency, and a cloud node may be needed for the full capability of a foundation model. The system may also activate a capability, reuse cached state, or route through multiple stages when split execution is beneficial. Thus, service delivery is jointly determined by capability availability, deployment state, network conditions, node resource status, and policy constraints. These considerations motivate the architectural mechanisms described in the next section.

\section{IDN Architecture}

IDN organizes intelligence delivery around six architectural functions: intelligence capability abstraction, compute resource integration, demand-driven capability deployment, capability-aware service routing, state-aware caching, and security and trust management, as shown in Fig. \ref{fig:idn_architecture}. These functions are tightly coupled. Capability abstraction defines what can be delivered; resource integration determines where it can execute; deployment prepares the substrate on which routing operates; routing determines whether cached state is useful; and security constraints influence every stage of the service path. Table \ref{tab:abstractions} lists the representative architectural descriptors in IDN.

\begin{table*}[t]
\small
\centering
\caption{Representative architectural descriptors in IDN.}
\label{tab:abstractions}
\begin{tabularx}{\textwidth}{p{3cm} p{6.0cm} Y}
\toprule
\textbf{Descriptor} & \textbf{Representative fields} & \textbf{Architectural role} \\
\midrule
Request descriptor $Q$ & capability, quality target, policy constraints, affinity token, cost budget & Expresses what intelligence is requested and under which service constraints. \\
Capability descriptor $C$ & task, quality tier, response-time requirement, security label, resource requirement, lineage & Describes a deployable intelligence capability that can be advertised, placed, and matched to requests. \\
Resource profile $R$ & accelerator type, memory, storage, runtime, locality, load, trust level & Summarizes what a node or domain can execute and under which current conditions. \\
State descriptor $S$ & state type, compatibility hash, sharing scope, size, reuse statistics, migration cost & Describes reusable artifacts, prefixes, tensor states, or results for cache management and routing. \\
Execution receipt $E$ & selected plan, capability version, node attestations, cache usage, policy verdict & Records how a request was served for debugging, auditing, and compliance. \\
\bottomrule
\end{tabularx}
\end{table*}

\subsection{Intelligence Capability Abstraction}

\begin{figure}[t]
  \centering
  \includegraphics[width=0.85\linewidth]{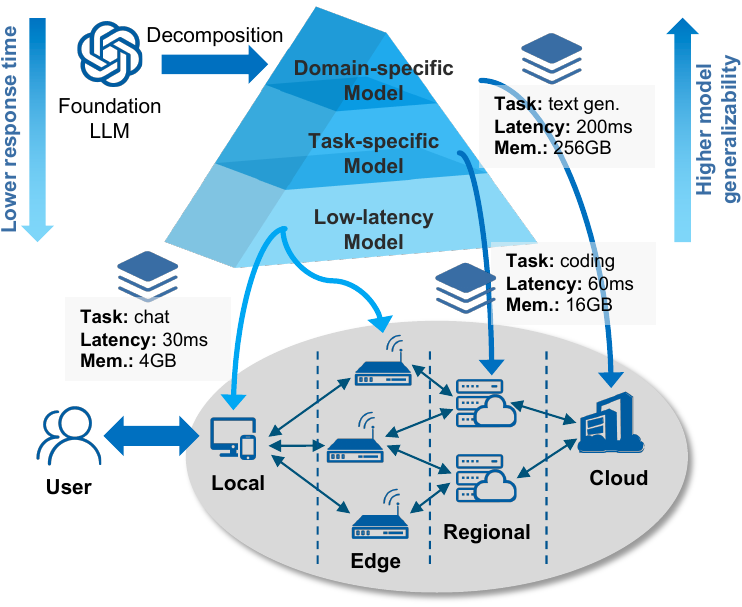}
  \caption{Capability abstraction and deployment in IDN. Foundation LLMs are decomposed into multiple capability variants, which are then placed on suitable computing nodes according to resource availability, service requirements, and policy constraints.}
  \Description{Capability abstraction and deployment in IDN. Foundation LLMs are decomposed into multiple capability variants, which are then placed on suitable computing nodes according to resource availability, service requirements, and policy constraints.}
  \label{fig:capability_abstract}
\end{figure}

A fundamental challenge for IDN is that the intelligence embodied in today's foundation models is not naturally aligned with the unit of network delivery. A large model may support many tasks, but as a deployment object it is too coarse-grained for efficient distribution. If intelligence were exposed only as a monolithic checkpoint, nodes would either need to replicate the same full model or give up serving that capability. This is inefficient in heterogeneous environments where requests differ in latency tolerance, accuracy targets, privacy requirements, and resource budgets. IDN therefore abstracts intelligence into smaller capability units that are meaningful for deployment, as shown in Fig. \ref{fig:capability_abstract}.

An \emph{intelligence capability} is a bounded and deployable unit of service competence derived from one or more foundation models. It does not preserve the full generality of the original model. Instead, it captures the subset of intelligence needed to support a class of requests under explicit quality, latency, security, and resource constraints. Such capabilities may be obtained through distillation, specialization, compression, adapter-based refinement, quantization, or other provider-specific mechanisms~\cite{hinton15distill,lora22,switch22,llmint8,smoothquant23}. The IDN architecture does not mandate a specific extraction method, only that the resulting unit is semantically meaningful, deployable, and stable enough to support placement, discovery, and routing.

IDN views capabilities at three levels. At the top level are \emph{capability classes}, such as translation, code assistance, visual recognition, or vertical-domain question answering. At the middle level are \emph{capability variants}, which encode tradeoffs among accuracy, latency, model size, privacy, and trust requirements. At the bottom level are \emph{deployable realizations}, such as distilled, quantized, or otherwise optimized model instances that implement a variant on a particular hardware type. This separation allows the network to reason about capability classes and variants while allowing local execution platforms to choose the concrete realization that best fits their resources.

We represent a capability by a descriptor
\begin{equation}
    C=\langle \texttt{name},\texttt{task},\texttt{quality},\texttt{latency},\texttt{security},\texttt{resource},\texttt{lineage}\rangle .
\end{equation}

\noindent Here, \texttt{name} denotes the capability class; \texttt{task} specifies the supported function or domain; \texttt{quality} records the expected service level, such as an accuracy tier or context limit; \texttt{latency} captures the response-time characteristic; \texttt{security} captures privacy and trust requirements; \texttt{resource} summarizes compute, memory, and storage requirements; and \texttt{lineage} records how the capability is derived from its originating model family. This descriptor provides the minimal information needed for capability advertisement, deployment planning, request matching, and compatibility checking.

Granularity is a key design issue. If the unit is too coarse, deployment degenerates into whole-model replication. If it is too fine, the control overhead of naming, advertising, and composing units may dominate any reuse benefit. IDN therefore adopts a \emph{service-effective granularity}, \ie a capability should be small enough to enable differentiated deployment across heterogeneous nodes, yet large enough to remain stable and meaningful as a service object. The granularity should also be demand-driven. Recurring or high-demand service patterns provide a natural basis for extracting and packaging capabilities because they identify the parts of intelligence most worth placing and reusing in the network. This abstraction makes hierarchical delivery possible. Once intelligence is decomposed into capability units, the network can advertise, deploy, route, and reuse them independently.

\subsection{Compute Resource Integration}

\begin{figure}[t]
  \centering
  \includegraphics[width=0.97\columnwidth]{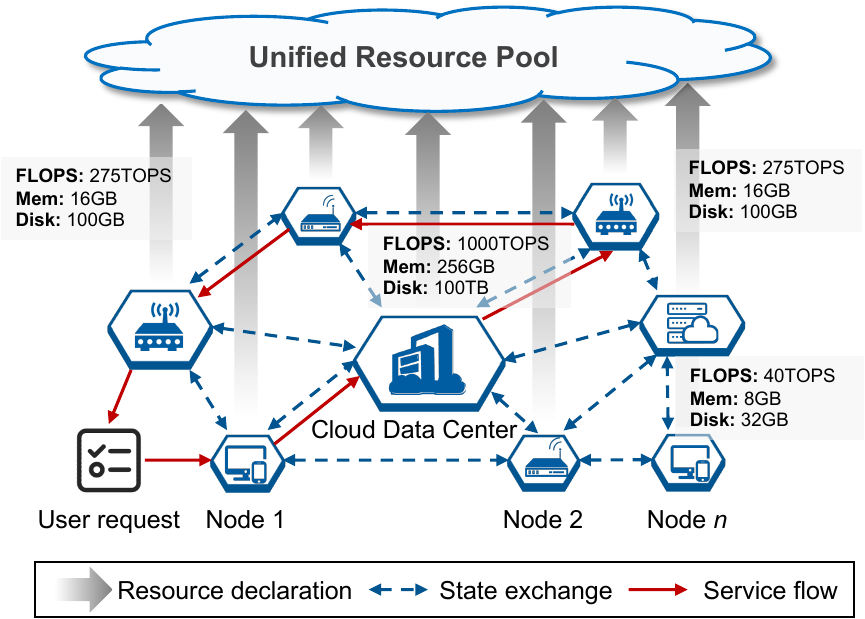}
  \caption{Unified compute resource pool in IDN. Compute nodes register with the resource pool by advertising their available compute resources and continuously reporting dynamic state information for capability-aware request routing.}
  \Description{Unified compute resource pool in IDN. Compute nodes register with the resource pool by advertising their available compute resources and continuously reporting dynamic state information for capability-aware request routing.}
  \label{fig:resource_integration}
\end{figure}

Distributed compute nodes must be incorporated into IDN as manageable suppliers of execution capability, as shown in Fig. \ref{fig:resource_integration}. We adopt a hierarchical integration model rather than direct client-side selection of arbitrary nodes. Similar to CDN operations, nodes are admitted, described, and monitored by the control plane of their administrative domain, while requests are steered using aggregated capability views rather than raw node-level exposure. This design improves stability and scalability because fast-changing node details do not need to be globally visible.

A compute node joins IDN by registering with a domain controller or broker responsible for local resource management. The registration creates a resource profile
\begin{equation}
R=\langle \texttt{hardware},\texttt{runtime},\texttt{capacity},\texttt{state},\texttt{locality},\texttt{trust}\rangle,
\end{equation}
where \texttt{hardware} captures accelerator type and memory capacity, \texttt{runtime} captures supported serving environments, \texttt{capacity} summarizes static resource limits, \texttt{state} summarizes dynamic load and model residency, \texttt{locality} expresses network and geographic position, and \texttt{trust} records identity and attestation properties. The profile is service-oriented. It exposes what the node can provide to the scheduler, not every hardware detail.

Because resource conditions change quickly, integration is not a one-time registration step. Local agents should continuously sense health, queueing condition, available memory, model residency, recent performance, and relevant network measurements. Detailed telemetry remains inside the local domain, while coarse capability summaries are propagated upward or across domains. This hierarchical sensing mirrors the fast local control loops and slower inter-domain coordination in~\cite{catsFramework,catsUsecases}. Once nodes are represented through a common resource view, IDN can support higher-level coordination, including staged execution, multi-node service realization, and adaptive deployment, without requiring a single global optimizer.

Resource integration must also be coupled with trust. A node that advertises resources should present verifiable identity and policy-relevant properties before serving protected requests or hosting sensitive capabilities. In this sense, integration is the mechanism by which heterogeneous compute resources become visible, manageable, and trustworthy service suppliers.

\subsection{Demand-Driven Capability Deployment}

After capabilities have been abstracted and resources integrated, IDN must decide which capabilities should be placed at which nodes. The objective is not to replicate all capabilities everywhere, but to position the right capability at the right level of the hierarchy. A capability placed too far from demand may incur unnecessary latency and bandwidth cost, while a capability placed too aggressively at constrained nodes may waste scarce memory or reduce system-wide efficiency.

Deployment in IDN is demand-driven. User demand is dynamic, unevenly distributed, and often concentrated around recurring service patterns. Therefore, the system should decide not only where existing capabilities are placed, but also when new variants should be constructed or activated. If a task pattern becomes frequent in a region, the system may derive a compact specialization, push it to suitable regional or edge nodes, and later withdraw it when demand subsides. This links capability abstraction to deployment, \ie the units worth creating are often the ones with enough reuse value under observed demand.

A useful high-level formulation is a constrained placement problem. Let $x_{c,n}\in\{0,1\}$ denote whether capability $c$ is placed at node or domain $n$. A deployment controller can be viewed as minimizing
\begin{equation}
\begin{aligned}
\min_{x} \quad &\sum_{q\in\mathcal{Q}} L(q,x)
+ \lambda C_{\mathrm{deploy}}(x)
+ \mu C_{\mathrm{net}}(x) \\
&+ \nu C_{\mathrm{risk}}(x) \\
\text{s.t.}\quad & \sum_c x_{c,n} m_c \le M_n, \quad
x_{c,n}\in\{0,1\}.
\end{aligned}
\end{equation}
Here, $L(q,x)$ captures the service latency or loss for request $q$ under placement $x$, $C_{\mathrm{deploy}}$ captures model loading and storage cost, $C_{\mathrm{net}}$ captures transfer or egress cost, $C_{\mathrm{risk}}$ captures policy and trust exposure, $m_c$ is the memory footprint of capability $c$, and $M_n$ is the resource budget of node $n$. The exact solver is provider-specific. The key point is that deployment must jointly consider demand locality, resource constraints, network cost, and policy risk.

Deployment must also be adaptive. Demand surges, resource fluctuations, node failures, and policy changes can invalidate a previously effective placement. IDN therefore treats deployment as a closed-loop process. It senses demand and infrastructure dynamics, places capabilities accordingly, and revises placement through migration, scale-out, scale-back, and rollback when conditions change. Existing model-serving systems already optimize placement and scheduling within a provider boundary~\cite{infaas21,cocktail22,alpaserve23,serverlessllm24,llumnix24}, but IDN exposes the interfaces to extend such decisions across a distributed network substrate.

\subsection{Capability-Aware Service Routing}

\begin{figure}[t]
  \centering
  \includegraphics[width=0.9\columnwidth]{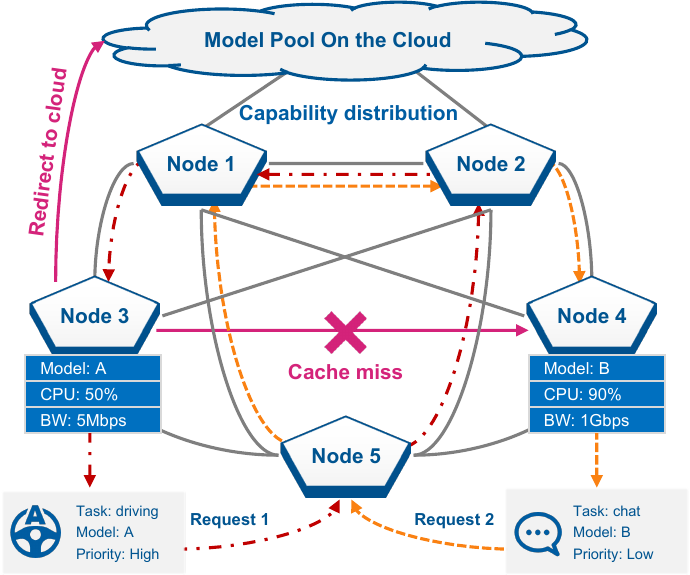}
  \caption{Service routing in IDN. IDN selects suitable compute nodes by jointly considering task requirements, deployed node capabilities, current resource states, and policy constraints.}
  \Description{Service routing in IDN. IDN selects suitable compute nodes by jointly considering task requirements, deployed node capabilities, current resource states, and policy constraints.}
  \label{fig:service_routing}
\end{figure}

Once capabilities are deployed, IDN must decide how an incoming request should be realized. Unlike conventional routing, the goal is not merely to forward a packet to a destination, but to select the execution location, capability variant, and scheduling decision that together satisfy the request. Some requests require stronger models or stricter quality guarantees, while others can be served by lightweight variants with lower latency or cost. Service routing is therefore capability-aware by design, as shown in Fig. \ref{fig:service_routing}.

We represent a request as
\begin{equation}
Q=\langle c,q,p,a,b\rangle,
\end{equation}
where $c$ denotes the requested capability, $q$ captures the target service quality, $p$ specifies policy or privacy constraints, $a$ is an optional state-affinity token, and $b$ is an optional resource or cost budget. Given $Q$, the scheduler determines the feasible set of execution plans
\begin{equation}
\Pi(Q)=\{\pi \mid \pi \text{ satisfies } c,q,p,b\},
\end{equation}
where a plan $\pi$ may correspond to a single node or a multi-stage path across nodes. IDN selects
\begin{equation}
\pi^*(Q)=\arg\min_{\pi\in\Pi(Q)} J(\pi,Q),
\end{equation}
with
\begin{equation}
\begin{aligned}
J(\pi,Q)=\;& \alpha T_{\mathrm{net}}(\pi)
+\beta T_{\mathrm{queue}}(\pi)
+\gamma T_{\mathrm{exec}}(\pi) \\
&+\delta T_{\mathrm{state}}(\pi,a)
+\epsilon C_{\mathrm{load}}(\pi)
+\zeta P_{\mathrm{policy}}(\pi,Q).
\end{aligned}
\end{equation}
Here, $T_{\mathrm{net}}$ denotes network transfer delay, $T_{\mathrm{queue}}$ queueing delay, $T_{\mathrm{exec}}$ execution time, $T_{\mathrm{state}}$ state transfer or reconstruction cost, $C_{\mathrm{load}}$ a load-balancing penalty, and $P_{\mathrm{policy}}$ a policy or trust penalty. The weights are deployment-specific. The key is that routing, scheduling, and policy enforcement are correlated with each other for AI services.

State-awareness is especially important. Interactive sessions, long-context inference, and split execution accumulate valuable intermediate state, including reusable prefixes and KV caches. If execution switches nodes, the system may need to move, reconstruct, or discard this state, which can dominate the latency gain from a shorter path. IDN therefore treats state affinity as a key routing consideration: the scheduler should preserve locality when reuse is valuable, but gradually divert traffic when queueing delay, contention, or resource pressure dominates. This is consistent with recent LLM-serving work showing that prefill and decode phases, KV-cache management, and state movement strongly affect tail latency and goodput~\cite{orca22,pagedattention23,sarathi24,distserve24,splitwise24,infinigen24,fairness24}.


Routing must also avoid hotspot formation. Demand, node load, and network conditions vary over time, and naive affinity-based routing may overload popular nodes. The load penalty in $J(\pi,Q)$ provides an architectural handle for balancing short-term locality against long-term service stability. Under overload, the system may route to a compatible warm instance, select a cheaper capability variant, reduce the admissible context window, or reject requests early rather than allowing unbounded queue growth. Thus, IDN routing is a service-quality control loop, not only a path-selection mechanism.

\subsection{State-Aware Caching}

Caching is essential for improving the efficiency and service quality of distributed AI systems. Unlike conventional content caching, AI-serving caches contain heterogeneous objects with different correctness, privacy, and mobility constraints. IDN therefore treats caching as state management rather than simple object reuse. The goal is to decide what state should be cached, where it should be placed, how it should be shared, and when it should be migrated or invalidated.

We distinguish four classes of cacheable objects. \emph{Artifact caches} store relatively stable deployment objects such as model weights, adapters, quantized variants, and runtime packages. \emph{Prefix caches} store reusable prompt modules, system instructions, retrieved-context templates, or other common input prefixes. \emph{Tensor-state caches} store intermediate execution state, such as KV caches or other activations, that can accelerate subsequent inference. \emph{Result caches} store reusable outputs for identical or semantically equivalent inputs under compatible model, decoding, and policy settings. They may include full responses when execution is deterministic or constrained, as well as deterministic subcomputations such as embeddings, retrieval outputs, or tool results~\cite{promptcache24,cachegen24,pagedattention23}. Orthogonal to type, each cached state has a sharing scope: public, tenant-shared, session-private, or hardware-bound.

A cached state object is described by a state descriptor containing its type, compatibility hash, size, sharing scope, reuse statistics, privacy label, decoding configuration when applicable, and migration cost. Cache admission can be expressed as
\begin{equation}
\mathrm{Benefit}(x,Q)=P_{\mathrm{hit}}(x,Q)\Delta L
-C_{\mathrm{transfer}}(x)-C_{\mathrm{storage}}(x)-C_{\mathrm{privacy}}(x),
\end{equation}
where $P_{\mathrm{hit}}(x,Q)$ is the predicted reuse probability of state $x$ for request class $Q$, $\Delta L$ is the expected latency reduction, and the remaining terms capture transfer, storage, and privacy costs. This formulation highlights that caching cannot be designed independently from routing and deployment. A cache is valuable only if future requests are likely to be routed to locations where it can be reused safely and efficiently.

IDN also supports cooperative caching across nodes. In generative AI workloads, state may be too large or too expensive to move as a single object. Prefill-decode separation, for example, may let compute-rich nodes perform prefill while latency- or memory-optimized nodes perform decode. The associated KV state must then be transferred, compressed, partitioned, or selectively retained along the execution path~\cite{distserve24,splitwise24,cachegen24}. Network awareness is necessary because reuse cost depends on bandwidth, congestion, path latency, and the communication pattern required to access or move state. The IDN architecture requires that cache policy must jointly consider model structure, compute resources, network conditions, routing decisions, and service objectives.

\subsection{Security and Trust Management}

Security in IDN must be considered in the context of distributed and collaborative inference. A service request may involve capabilities derived from different models, nodes operated by different domains, cached states produced by previous requests, and intermediate results exchanged across the network. Protecting only the transport channel is therefore insufficient. IDN must provide architectural support for ensuring that data is handled under appropriate constraints, model components are trustworthy, and inference results can be associated with a reliable execution process.

A key principle is to make trust visible to the architecture. Compute nodes should expose verifiable identities and security-relevant properties before serving protected requests. Similarly, intelligence capabilities should carry authenticated metadata about origin, version, dependency chain, and admissible policy scope. This allows the system to reason not only about whether a node has enough resources, but also about whether it is authorized and trusted to execute a particular request. In multi-domain deployments, such information is necessary for coordinating authentication and authorization across operators with different policies and trust assumptions.

Security constraints should participate in deployment and routing. Different requests impose different privacy, locality, and isolation requirements. A latency-critical request without strict locality constraints may be routed to a nearby lightweight capability, whereas a request involving private or regulated data may require local execution, stronger isolation, or a node within a specific administrative domain. These choices trade off privacy, performance, and resource efficiency. IDN therefore treats security and privacy as scheduling constraints rather than after-the-fact application checks. Trusted execution environments, confidential VMs, and verifiable inference techniques provide useful building blocks, but they should be exposed through capability and resource descriptors rather than hidden inside provider-specific implementations~\cite{scone16,sanctum16,sevSnp20,slalom19,delphi20,opaque17}.

IDN must also account for the full lifecycle of intelligence delivery. Model components, adapters, runtime packages, cached states, and execution environments can all become attack surfaces. Model backdoors and poisoned components show that the ML supply chain itself can be compromised~\cite{gu17badnets,liu18trojan}. The architecture should therefore support provenance tracking, policy-based access control, runtime monitoring, and invalidation or rollback of unsafe components. Execution receipts can record capability versions, node attestations, cache usage, and policy decisions for auditing and incident response. The goal is not to mandate a single defense mechanism, but to provide common hooks for maintaining trust in distributed inference.

\section{IDN Architecture Development}

The previous sections describe the core architectural mechanisms of IDN. A natural next question is how such an architecture can be developed, validated, and deployed. We view IDN as an architecture that should evolve through application-driven design, prototype systems, experimental testbeds, and incremental deployment.

\subsection{Application-Driven Design}

IDN should be developed toward concrete AI applications. Different applications stress different parts of the architecture. Interactive assistants and copilots emphasize low latency, session continuity, and state reuse. Real-time translation and AR/VR applications require predictable response time and close placement of lightweight capabilities. Enterprise retrieval-augmented generation stresses policy compliance, locality of private data, and controlled reuse of retrieved context~\cite{lewis20rag}. Robotics and multimodal perception introduce stronger requirements on reliability, edge execution, and graceful degradation.

These applications are design drivers as well as evaluation workloads. They help determine the right granularity of capability abstraction, the fields exposed in descriptors, the resource states that must be advertised, and the security constraints that should influence deployment and routing. In this sense, application development provides feedback into the architecture by revealing where provider-local model serving is insufficient for network-level intelligence delivery.

\subsection{Prototype System and Testbed}

A practical IDN prototype should separate architectural functions from provider-specific serving implementations. At a high level, such a prototype would include a capability registry, resource broker, deployment controller, service router, state/cache manager, and trust manager. These components can be built as an overlay on existing infrastructure such as Kubernetes, service meshes, Ray, GPU inference runtimes, and model-serving frameworks~\cite{moritz18ray,kubernetes,clipper17,orca22,pagedattention23}. IDN need not replace these systems. Rather, it exposes capability, resource, state, and policy information across them.

A useful testbed should span cloud clusters, regional nodes, edge servers, and local or on-premise nodes. Such a testbed would allow researchers to study cross-layer questions that single-cluster inference benchmarks cannot capture, including capability placement, state migration, load-aware routing, privacy-aware scheduling, failure recovery, and multi-domain trust.

\subsection{Evaluation and Incremental Deployment}

Evaluating IDN requires metrics beyond network throughput or single-cluster inference speed. From the user perspective, the metrics include time-to-first-token, time-per-output-token, tail latency, completion rate, and degradation under overload. From the system perspective, IDN should be evaluated by accelerator utilization, memory pressure, model loading overhead, cache hit ratio, state-transfer cost, and load-balancing effectiveness. From the network perspective, relevant metrics include wide-area traffic reduction, inter-domain transfer cost, and sensitivity to congestion or failures.

The deployment of IDN shall proceed incrementally. The first stage may occur within a single provider domain that already controls cloud and edge resources. The next stage is integration with CDN-like and edge infrastructures, extending their control planes from content objects to intelligence capabilities and reusable state. Cross-domain intelligence delivery requires stronger standardization of capability descriptions, resource summaries, state descriptors, and execution evidence. Therefore, it requires gradually standardizing the interface needed to discover, deploy, route, reuse, and trust distributed AI services.

\section{Open Questions}

IDN raises several research questions. A central issue is the abstraction boundary exposed to the network. This paper argues for deployable capabilities and associated descriptors because they make placement, routing, and state reuse visible to the architecture. However, future AI services may involve multi-agent workflows, tool-using pipelines, or dynamically composed reasoning processes. The appropriate network-visible unit for such services may be richer than a single capability.

Another challenge is the economic and operational model of cross-domain intelligence delivery. CDNs and cloud platforms have historically hidden settlement, peering, and placement logic behind provider-specific contracts. IDN makes the situation more complex because execution capacity, cached state, model placement, and policy compliance may all carry economic value. If intelligence can be placed and served across domains, the architecture may require explicit mechanisms for accounting, incentive alignment, and policy negotiation.

State mobility is also underexplored. Distributed inference creates many forms of reusable state, including prompt prefixes, KV caches, intermediate activations, retrieval results, and session context. Moving such state can reduce repeated computation, but it may consume bandwidth, increase latency, violate compatibility constraints, or weaken privacy guarantees. The question of when state should move, be recomputed, or be discarded is both a system problem and a networking problem.

Trust and verification introduce another set of open problems. Provenance, signed capability metadata, and node attestation can establish where a request executed and which components were involved, but they do not prove semantic correctness or eliminate all side channels. Practical IDN deployments will need combinations of attestation, runtime monitoring, selective verification, privacy-preserving inference, and auditing~\cite{slalom19,delphi20,opaque17}.

Finally, progress on IDN will require shared traces and benchmarks. Without workloads that capture model popularity, request locality, context reuse, accelerator heterogeneity, privacy constraints, and multi-domain operation, it will be difficult to compare placement algorithms, routing policies, caching strategies, or fairness mechanisms. Building this empirical foundation may be as important as designing any single control algorithm.

\section{Looking Forward}

IDN is not intended to replace today's Internet or centralized AI clouds. Rather, it identifies a path for extending the Internet so that distributed AI services can be delivered more efficiently, reliably, and securely. As AI workloads become increasingly interactive, personalized, multimodal, and privacy-sensitive, the network must reason not only about reachability, but also about capability, resource availability, reusable state, and policy constraints.
This paper presented IDN as an architectural framework for this transition. Its key idea is to make intelligence a deliverable network service through the six interdependent mechanisms.

Looking forward, IDN should be viewed as the next generation of Internet architecture for the AI age, which defines the abstractions and interfaces needed for the Internet to actively participate in delivering intelligence, rather than merely transporting requests to remote AI services.

\balance
\bibliographystyle{ACM-Reference-Format}
\bibliography{references}

\end{document}